\documentclass[conference]{IEEEtran}
\IEEEoverridecommandlockouts
\usepackage{cite}
\usepackage{longtable}
\usepackage{booktabs} 
\usepackage{amsmath}
\usepackage{mathtools}
\usepackage{algorithm, tabularx}          
\usepackage{algpseudocode}      
\usepackage{amssymb}
\usepackage{multirow}
\usepackage{array}
\usepackage{makecell}
\usepackage{float}
\usepackage{chngcntr}
\usepackage{hhline}
\usepackage{kotex}
\usepackage{xcolor}
\usepackage{enumerate}
\usepackage[shortlabels]{enumitem}
\usepackage{amsmath}
\usepackage{hyperref}

\DeclareMathOperator*{\argmin}{arg\,min}
\DeclarePairedDelimiter\floor{\lfloor}{\rfloor}

\def\BibTeX{{\rm B\kern-.05em{\sc i\kern-.025em b}\kern-.08em
    T\kern-.1667em\lower.7ex\hbox{E}\kern-.125emX}}

\newcommand{\ours}{CITIES}

\begin{document}

\title{CITIES: Contextual Inference of Tail-item Embeddings for Sequential Recommendation}
\author{\IEEEauthorblockN{Seongwon Jang, Hoyeop Lee, Hyunsouk Cho, Sehee Chung}
\IEEEauthorblockA{Knowledge AI Lab., NCSOFT Co., South Korea\\
\{swjang90, hoyeoplee, dakgalbi, seheechung\}@ncsoft.com}
}
\maketitle

\begin{abstract}
Sequential recommendation techniques provide users with product recommendations fitting their current preferences by handling dynamic user preferences over time. Previous studies have focused on modeling sequential dynamics without much regard to which of the best-selling products (i.e., head items) or niche products (i.e., tail items) should be recommended. We scrutinize the structural reason for why tail items are barely served in the current sequential recommendation model, which consists of an item-embedding layer, a sequence-modeling layer, and a recommendation layer. Well-designed sequence-modeling and recommendation layers are expected to naturally learn suitable item embeddings. However, tail items are likely to fall short of this expectation because the current model structure is not suitable for learning high-quality embeddings with insufficient data. Thus, tail items are rarely recommended. To eliminate this issue, we propose a framework called \ours, which aims to enhance the quality of the tail-item embeddings by training an embedding-inference function using multiple contextual head items so that the recommendation performance improves for not only the tail items but also for the head items. Moreover, our framework can infer new-item embeddings without an additional learning process. Extensive experiments on two real-world datasets show that applying \ours ~to the state-of-the-art methods improves recommendation performance for both tail and head items. We conduct an additional experiment to verify that \ours ~can infer suitable new-item embeddings as well.
\end{abstract}


\section{Introduction}
The objective of a sequential recommendation task is to recommend products that users would like to have, given their historical behaviors as a sequence. In many real-world applications, users' interests inherently evolve, influenced by sequential behaviors. To cope with this dynamic situation, various methods to capture the sequential dynamics from users' action history have been proposed~\cite{markov1, markov2, gru4rec, carnn, sasrec, caser, bert4rec}. These methods can effectively model the sequential dynamics. However, most existing models have not been properly tested in terms of long-tail recommendation, because items having less than a threshold number of actions from users are normally discarded via the preprocessing procedure. Anderson~\cite{anderson} introduced the term, \textit{long tail}, to describe the phenomenon by which niche products can grow to become a large share of total sales. According to the results of his analyses, companies such as Amazon, which apply \textit{the long-tail} effect, successfully earn most of their revenue not from the best-selling products (i.e., head items), but niche products (i.e., tail items). Although tail items are important, it is well-known that general recommendation models (e.g., matrix factorization~\cite{mf}) barely serve tail items because of the skewed distribution of product popularities~\cite{longtail1, longtail2, longtail3}. Models trained using skewed distributed data are prone to a popularity bias. As a result, they recommend head items over tail or new items, even if the latter would be viewed favorably.

This phenomenon also occurs in sequential recommendation models because they are trained by the skewed distributed data. Moreover, when the user behavior sequence contains tail items and their embeddings are poorly trained because of the insufficient training data, sequential recommendation models may struggle to understand user behavior. Recently, a method of treating popularity bias in the context of sequential recommendation was proposed. Kim et al.~\cite{sdiv} proposed S-DIV, which is based on a gated recurrent unit (GRU)~\cite{gru}. They attempted to recommend tail items more frequently by replacing each tail item with a cluster of them based on content features. Their approach had a limitation in which the embedding of tail items could not be directly obtained.

In this study, we scrutinize the structural reason for why the tail items are barely served compared with the head items in general sequential recommendation models comprising an item-embedding layer, a sequence-modeling layer, and a recommendation layer. All existing sequential recommendation methods lack efforts to explicitly improve tail-item embeddings. Item embedding plays a major role in the model because it is used twice: as input to the sequence-modeling layer to represent the meaning of each item, and as input to the recommendation layer to rank items in the order in which a user will likely interact. As a result, the item embeddings greatly affect the model performance. Previous studies focused on architecture designs of the sequence-modeling layer and expected that well-designed architecture would naturally learn suitable item embeddings via back-propagation. However, unlike the head item, the tail item is likely to fall short of the expectation because the general model structure is not suitable for learning high-quality embedding with insufficient data. Thus, we require an alternative that will enable us to overcome this situation and obtain suitable tail-item embeddings.

Inspired by some remarkable natural language processing (NLP) studies~\cite{cw1, cw2, cw3, cw4} that allow us to obtain high-quality embeddings of rare words with their contextual information, we propose the \textbf{C}ontextual \textbf{I}nference of \textbf{T}ail-\textbf{i}tem \textbf{E}mbedding\textbf{s} (\textbf{\ours}) framework, which is easily compatible with extant sequential recommendation models. \ours ~aims to enhance the quality of tail-item embeddings using contextual items so that the performance of sequential recommendation improves for both tail and head items. We call a set of nearby items the contextual items in the sequential recommendation. Because any item can be consumed by multiple users, the item might have multiple contextual items. Leveraging this trait, we propose an embedding-inference function in our framework to determine the tail-item embeddings using the multiple contextual items. The embedding-inference function is trained to take the multiple contextual items of the head item as input and reproduce the head-item embeddings pre-trained with the general sequential recommendation model. With the intent to train the function with sufficient high-quality data, we utilize only the head item as the reproducing target. After training, we apply the function to the tail items to enhance their embeddings.

The enhanced tail-item embeddings directly affect the recommendation layer and the item-embedding layer. As a result, the tail items are ranked higher when users are more likely to interact with them. Performance improvement of the head items is achieved when the user behavior sequence contains the tail items. Low-quality tail-item embeddings in the sequence hinder the neural architectures that model the sequential dynamics. On the other hand, the enhanced tail-item embeddings help the neural architectures to model sequential dynamics. Moreover, using \ours, when new items appear, it can also infer their embeddings without further learning. If a new item is consumed more than once, we can extract its contextual items. Using the contextual items, we can then infer the embedding of the new items in the same way as that of the tail items. To verify our expectations, we conduct extensive experiments on real-world datasets from Yelp and Amazon.

Our study provides three main contributions. First, we propose an easily compatible framework, \ours, that infers the tail-item embeddings using an embedding-inference function learned from head-item embeddings to reduce the popularity bias in sequential recommendation. Second, \ours ~can infer embeddings of the new items that did not exist during training without another learning process. Third, we show quantitatively that applying our framework improves state-of-the-art sequential recommendation methods on two real-world datasets for both tail and head items. Moreover, we conduct additional experiments to analyze the contributions of each component and the impact of factors in the proposed framework. 
\section{Related work}
\subsection{Sequential Recommendation}
Some early works on sequential recommenders captured sequential dynamics from user behaviors using Markov chains (MC)~\cite{markov1, markov2} and assumed that the user’s next action would be highly related to their last (or last few) action(s). Recently, inspired by sequence learning in NLP, several methods based on deep neural networks have been proposed to learn sequential dynamics. Recurrent neural network (RNN)-based methods~\cite{ carnn, gru4rec, gru4rec+} have been increasingly used to model sequential dynamics, wherein the hidden states of the RNN reflect a summary of the sequences. These methods allow for long-term semantics of the sequence to be uncovered. Other than RNN-based methods, various methods based on convolutional neural networks (CNN) and Transformer~\cite{transformer} have been used to provide sequential recommendations. Tang and Wang~\cite{caser} utilized convolutional filters to capture short-term contexts by regarding sequential features as local features of an image. Although effective, their mechanism had limitations in that it reflected long-term interests. Kang and McAuley~\cite{sasrec} proposed a Transformer-based model to represent the user's interests from not only all behaviors in the sequences, such as RNNs, but also for a few behaviors, such as MCs and CNNs. By improving their model, Sun et al.~\cite{bert4rec} proposed a bidirectional sequential model called BERT4Rec, which effectively trained the Transformer by predicting randomly masked items in the sequence. However, none of these methods considered the popularity bias, which tends to only recommend head items.

\subsection{Long-tail Recommendation}
In the field of general recommender systems, studies that well-emphasize tail items can be divided into two directions: post-processing and learning-to-rank. The first direction heuristically re-ranks the pre-ranked item list considering popularity bias. Adomavicius and Kwon~\cite{reranking1} re-ranked a pre-ranked item list in ascending order of popularity. Antikacioglu and Ravi~\cite{reranking2} formulated reranking as a subgraph selection problem from a super-graph of the pre-ranked items to select those by explicitly considering both the accuracy and the popularity bias. Although this direction can be applied regardless of the type of the recommendation model, it is often suboptimal because of the heuristic nature of the method. The second direction mitigates this problem by directly training the model by considering both accuracy and popularity bias. Cheng et al.~\cite{ltr1} proposed a method that directly learned a ranking function that imposed high values on tail and head items. Zou et al.~\cite{ltr2} improved their method using reinforcement learning and by avoiding the local optimal solution by learning the optimal global policy. However, these methods were learned based on the collaborative filtering algorithm~\cite{mf}. Thus, sequential dynamics were ignored. Recently, Kim et al.~\cite{sdiv} treated the popularity bias in the context of sequential recommendations by clustering and relocating consumed tail items to provide a pseudo ground truth, thereby allowing a GRU to directly learn the ranking function. Although effective, their approach was applied only to a GRU, making it difficult to apply the approach to other neural networks. However, \ours ~can be easily applied to other neural networks because it simply updates the tail-item embedding.
\section{Preliminary} 
Let $\mathcal{U}=\left\{u_{1}, \ldots, u_{|\mathcal{U}|}\right\}$ denote a set of users, $\mathcal{I}=\left\{i_{1}, \ldots, i_{|\mathcal{I}|}\right\}$ be a set of items, and ${S}^{u}=\left(s_{L-\ell+1}^{u}, \ldots, s_{L}^{u}\right)$ denote an item sequence ($s_{L}^{u} \in \mathcal{I}$). 
${S}^{u}$ represents $\ell$ recent items consumed by the user, $u$, and $L$ denotes the current time step.
The sequence can be obtained using all possible user actions, such as purchasing, writing a review, and browsing. Given the item sequence, ${S}^{u}$, the sequential recommendation model aims to recommend the item that user $u$ will interact during the next time step.

As mentioned previously, the sequential recommendation model has three types of layers, as shown in Figure~\ref{fig:ours} with black boxes and red arrows - the item-embedding layer $f_{\theta_{E}}$, the sequence-modeling layer $f_{\theta_{M}}$, and the recommendation layer $f_{\theta_{R}}$.
First, the item-embedding layer maps an item to a dense lower-dimensional vector representation.
Second, the sequence-modeling layer captures the sequential dynamics in the sequence of items which are encoded via an item-embedding layer. 
Lastly, the recommendation layer compares the similarity between the user's status representation produced via the sequence-modeling layer and the item embeddings and computes a probability vector of items with which the user, $u$, with the sequence, $S^u$, will interact with next.

\subsection{Item-embedding Layer} 
In a sequential recommendation model, the input is the recent $\ell$ items (i.e., ${S}^u$) in which each item is represented by a unique index. Generally, the item-embedding layer, $f_{\theta_{E}}$, maps each item index to a $d$-dimensional real-valued dense vector. $f_{\theta_{E}}$ operates in an element-wise manner for each item index of the input sequence.
The item-embedding layer can be designed in various ways. For example, the layer can be the embedding lookup matrix, which is similar to the neural language modeling process~\cite{nlm} of NLP. In some cases, to explicitly reflect an item's position in the input sequence, $f_{\theta_{E}}$ can include a positional embedding matrix. Some techniques, such as layer normalization~\cite{ln} and dropout~\cite{dropout}, can also be applied to effectively train the model.
In the item-embedding layer, the sequence of items is represented as
\begin{align}
\nonumber    E^{u} &= {f_{\theta_{E}}(S^u)}^\intercal \\ 
                     &= {f_{\theta_{E}}([ s_{L-\ell+1}^u , \cdots, s_{L}^u ])}^\intercal \\
\nonumber            &= [ f_{\theta_{E}}(s_{L-\ell+1}^u) , \cdots, f_{\theta_{E}}(s_{L}^u) ]^\intercal,
\end{align}
where $E^{u} \in \mathbb{R}^{\ell \times d}$ indicates the embedding vectors of items with which user $u$ has interacted. 

\subsection{Sequence-Modeling Layer} 
The sequence-modeling layer, $f_{\theta_{M}}$, captures the user's interest at the $L^{\textit{th}}$ time step by modeling the sequential dynamics in the sequence of the items.
The layer takes the sequence of the item-embedding vectors, $E^{u}$, as input and produces the current status, $m^{u}$, of user $u$ as follows:
\begin{equation}
m^{u} = f_{\theta_{M}}(E^{u}).
\end{equation}
For $f_{\theta_{M}}$, we can use any sequential deep neural architecture. Generally, among the architectures, RNNs and Transformer~\cite{transformer, bert} are the most popular, and the selecting algorithm depends on the sequential dynamics of the items. 
Using BERT4Rec~\cite{bert4rec} as an example, we can expand $f_{\theta_{M}}$ into equations (\ref{eq:bert}). BERT4Rec appends a special token, [mask], to the end of the sequence to extract the user's current state from the sequence shown in equation (\ref{eq:bert:1}). $N$-stacked self-attention layers in equations (\ref{eq:bert:2})--(\ref{eq:bert:3}) take $E^u$ as input and iteratively compute the hidden representation, $H^n$. The self-attention layer contains two sub-layers: multi-head attention ($\operatorname{MHA}$) and position-wise feed-forward ($\operatorname{PWFF}$). $\operatorname{MHA}$ models the dependencies between items by attending to different positions from multiple perspectives, and $\operatorname{PWFF}$ endows the model with nonlinearity and considers interactions between different latent dimensions. To effectively train the model, layer normalization (LN), dropout, and residual connection~\cite{residual} are applied between sub-layers. ${h}_{[mask]}^{N}$, the final hidden representation of the [mask] token after $N$ self-attention layers, is exploited to represent the sequence vector. $m^{u}$ is computed by applying a feed-forward layer with Gaussian error linear-unit ($\operatorname{GELU}$) activation~\cite{gelu} to ${h}_{[mask]}^{N}$ as follows:
\begin{subequations}
\label{eq:bert}
\begin{align}
\label{eq:bert:1} E^u &= \left[ f_{\theta_{E}}(s_{L-\ell+1}^u) , \cdots, f_{\theta_{E}}(s_{L}^u), f_{\theta_{E}}(s_{[mask]}) \right]^\intercal, \\
\label{eq:bert:2} {A}^{n-1}&=\operatorname{LN}\left({H}^{n-1}+\operatorname{Dropout}\left(\operatorname{MHA}\left({H}^{n-1}\right)\right)\right), \\
\nonumber H^n &=\operatorname{LN}\left({A}^{n-1}+\operatorname{Dropout}\left(\operatorname{PWFF}\left({A}^{n-1}\right)\right)\right) \\
\label{eq:bert:3} & \qquad \qquad \qquad \qquad \qquad \qquad \forall n=1,\dots,N, \\
\label{eq:bert:4}m^{u} &=\operatorname{GELU}\left({h}_{[mask]}^{N} {W}+{b}\right),
\end{align}
\end{subequations}
where $H^0=E^u$, $W$ is the weight matrix, and $b$ is the bias vector. For details on expanding $f_{\theta_{M}}$ into GRU4Rec, please refer to~\cite{gru4rec}.

\subsection{Recommendation Layer} 
To determine which items are relevant to user $u$, we feed the user's status vector, $m^{u}$, into the recommendation layer, $f_{\theta_{R}}$. The relevance score is computed as follows:
\begin{equation}
r^{u}=f_{\theta_{R}}\left(m^{u}, \theta_{E}\right),
\end{equation}
where $r^{u}\in \mathbb{R}^{1 \times |\mathcal{I}|}$ is the probability vector of whole items at the $(L+1)^{\textit{th}}$ time step.
As usual, to model $f_{\theta_{R}}$, we harness the classical matrix factorization term~\cite{mf}. That is, $r^{u}$ is computed from the inner product between the sequence vector, $m^{u}$, and each item-embedding vector. A bias term of the item can be added to the inner product value. Generally, the item-embedding vector is shared in the recommendation layer to alleviate overfitting. It is known that using shared item embeddings improve recommendation performance~\cite{gru4rec+, sasrec, bert4rec}.
\subsection{Model Training} 
We can train the sequential recommendation model using various loss functions, such as negative log-likelihood loss and Bayesian personalized ranking loss~\cite{bpr}. These loss functions are subtly different, but they commonly aim to increase the probability, $r_{i^*}^u$, of a ground-truth item, $i^*$. Regarding negative log-likelihood loss as an example, $f_{\theta}$ is optimized as follows:
\begin{equation}
\hat{\theta} = \argmin_{\theta} \sum_{u \in \mathcal{U}}\sum_{i^* \in {G}_{i^*}^{u}}-\log \frac{r_{i^*}^{u}}{\sum_{j \in \mathcal{I}}r_{j}^{u}},
\end{equation} 
where $\theta=\{\theta_E, \theta_M, \theta_R\}$, and ${G}_{i^*}^{u}$ is a set of the ground-truth items in $S^u$. Several regularization techniques, such as $l_2$ regularization, can be used for training to prevent $\hat{\theta}$ from overfitting.

${G}_{i}^{u}$ can be configured differently depending on recommendation models. For example, BERT4Rec~\cite{bert4rec} is trained by predicting the randomly masked items in the input sequence. Specifically, when $S^u=\left(s_{1}^{u}, s_{2}^{u}, s_{3}^{u}, s_{4}^{u}, s_{5}^{u} \right)$, $S^u$ can be converted to $\left(s_{1}^{u}, [mask], s_{3}^{u}, [mask], s_{5}^{u} \right)$. Thus, the set of ground-truth items ${G}_{i}^{u}=\{s_{2}^{u}, s_{4}^{u}\}$. The size of ${G}_{i^*}^{u}$ varies depending on how the sequence is utilized for training.
\section{Proposed Framework}
\begin{figure}
  \centering
  \includegraphics[width=84mm]{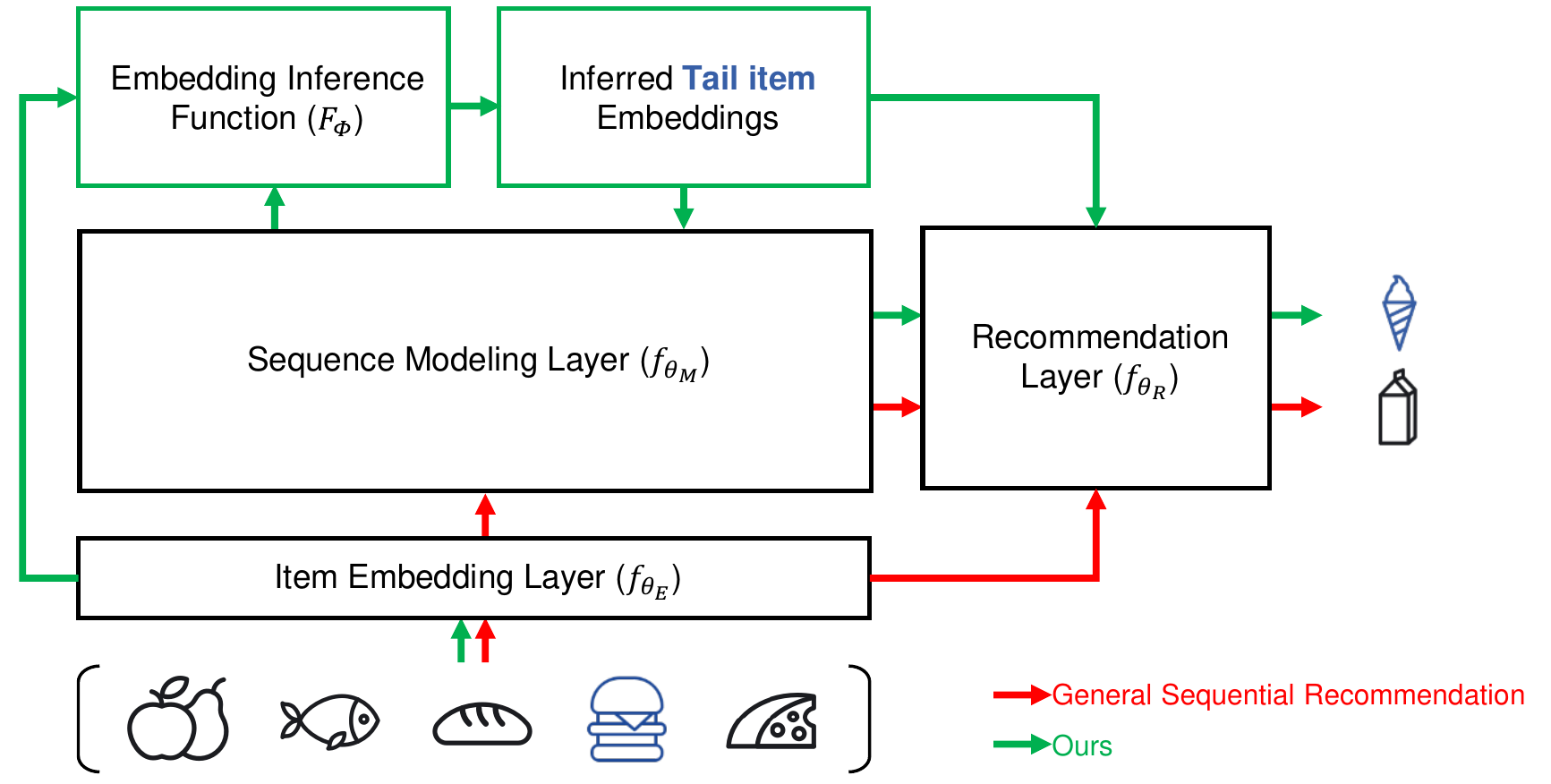}
  \caption{Comparison of the general sequential recommendation method and \ours. The black items belong to the head items, and the blue items indicate tail items.}
  \label{fig:ours}
\end{figure} 
In this section, we present \ours, whose framework is shown in Figure~\ref{fig:ours} with green boxes and arrows. We first provide a high-level overview of the proposed model. We then zoom into each of its phases.

\subsection{Framework Overview}

\ours ~consists of three phases. First, we pre-train the general sequential recommendation model. The pre-training is conducted as a general sequential recommendation learning method, as shown in Figure~\ref{fig:ours} with the red arrow. By pre-training, we obtain high-quality head-item embeddings used as training pairs for the embedding-inference function. Second, after dividing the items into head and tail groups based on their popularity in the training data, we construct training pairs (i.e., head-item embeddings and their contextual items) and train the embedding-inference function using the training pairs to reproduce the pre-trained embeddings of the head items from their set of contexts. To cope with the inference situation in which the tail item has only a few contexts, we formulate the task-inferring embeddings of the items from their contextual items as a few-shot learning task. In each iteration, the model is asked to predict the head item’s embedding with only $\kappa$ randomly sampled contexts of the item. Additionally, we leverage the pre-trained parameters of the sequence-modeling layer to effectively train the function. The training-completed function then infers the tail-item embeddings taking their contexts as input. Finally, relevance scores for each sequence are newly computed by applying inferred embeddings to the pre-trained model. 

\subsection{First Phase: Pre-training}
The goal of the first phase is to obtain the head item's high-quality embeddings and pre-trained parameters of the sequence-modeling layer, which are used to robustly train the function in the second phase. To obtain them, we adopt BERT4Rec~\cite{bert4rec} and GRU4Rec~\cite{gru4rec} in our framework, because they show the state-of-the-art performance and any model using the item-embedding layer can be used for pre-training. For this, we follow the model architectures and training procedures used in ~\cite{bert4rec, gru4rec}. BERT4Rec and GRU4Rec utilize Transformer and GRU, respectively, as their sequence-modeling layer, $f_{\theta_{M}}$. After pre-training, the parameters of $f_{\theta_{E}}$ and $f_{\theta_{M}}$ are sent to the next phase.

\subsection{Second Phase: Inferring Tail-item Embeddings}

\subsubsection{Problem Formulation}
The goal of the second phase is to improve embeddings for the tail items, $\mathcal{I}^t=\left\{i_{1}^t, \ldots, i_{|\mathcal{I}^t|}^t\right\}$. Because the tail items occur only a few times in the training data, it is difficult to directly learn their high-quality embeddings. 
Our solution is to learn the embedding-inference function, $F_\phi(\cdot)$, using only head items ($\mathcal{I}^h=\mathcal{I}\setminus \mathcal{I}^t$) that have relatively higher-quality embeddings than do the tail items as a target item. 
Let $C_{i} = \{C_{i}^1, \ldots, C_{i}^K\}$ be the context set of a target item, $i$, and $C_{i}^k = (s_{[i]-{\omega_1}}^k, \dots, s_{[i]}^k, \dots s_{[i]+{\omega_2}}^k)$ must be a subsequence of some user, $u$'s, sequence, ${S}^u$, for target item $i$, where $K$ is the size of the context set, $\omega_1$ and $\omega_2$ are window sizes of the left and right contextual items, respectively, and $[i]$ presents the order of item $i$ being consumed.
Our embedding-inference function, $F_\phi(\cdot)$, takes the context set, $C_{i}$, as input. 
During training, the target item should belong to the head items (i.e., $i\in\mathcal{I}^h$), and the function learns to generate an embedding vector close to its high-quality embedding vector, $f_{\theta_{E}}(s_{[i]})$, which was pre-trained during the first phase.
On the other hand, during testing, for the target item belonging the tail items (i.e., $i\in\mathcal{I}^t$), or new items (i.e., $i\notin\mathcal{I}$), the function estimates the item-embedding vector.

To cope with the inference situation in which the tail item has only a few contexts, we formulate a training procedure of $F_\phi(\cdot)$ as a few-shot learning task. In each iteration for target item $i$, we randomly sample only $\kappa$ contexts (if $\kappa>K$, $\kappa=K$) 
from $C^{i}$ to predict the embedding of $i$. This training scheme can simulate the inference situation such that $F_\phi(\cdot)$ can robustly infer embedding well, even when $K$ is few. The training objective for $F_\phi(\cdot)$ is to minimize the squared distance between the pre-trained embedding and inferred embedding vectors:
\begin{equation}
\hat{\phi}=\argmin_{\phi} \sum_{i\in\mathcal{I}^h} \sum_{C_{i}^{\kappa} \sim C_i} \left(F_{\phi}\left(C_{i}^{\kappa}\right)- f_{\theta_{E}}(s_{[i]})\right)^2,
\end{equation}
where $C_{i}^{\kappa} \sim C_{i}$ means that the $\kappa$ contexts are randomly sampled from the contexts set, $C_{i}$, containing the target head item, $i\in\mathcal{I}^h$. 
After $\hat{\phi}$ is trained, it infers the embeddings of the tail item, $i\in\mathcal{I}^t$, by taking all the contexts, $C_{i}$, as input.

\subsubsection{Embedding-Inference Function}
\begin{figure}
  \centering
  \includegraphics[width=84mm]{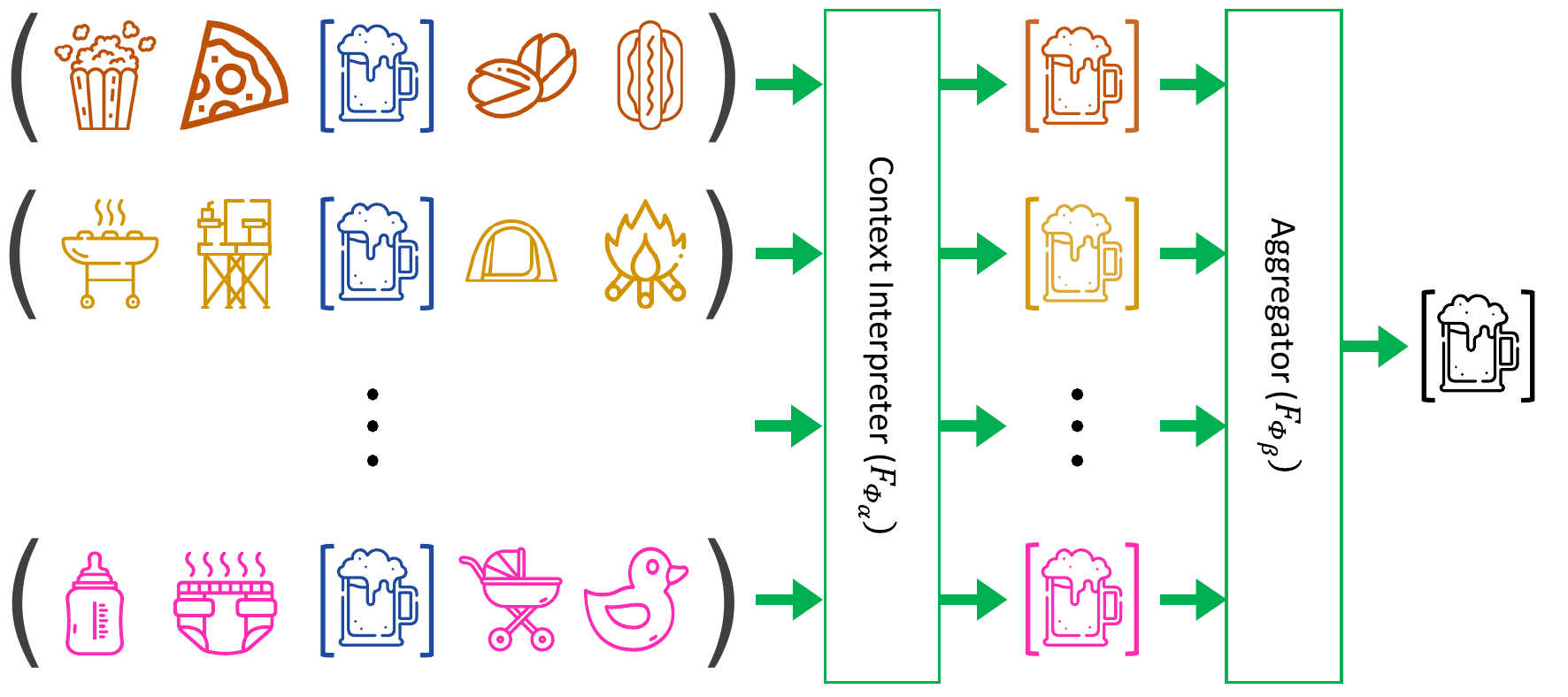}
  \caption{Embedding-inference function.}
  \label{fig:town}
\end{figure} 
Here, we present the details of the embedding-inference function, $F_\phi(\cdot)$, as shown in Figure~\ref{fig:town}. 
To understand the meaning of this item from multiple contextual items, $F_\phi(\cdot)$ should perform two types of roles: understanding the semantic of each context and determining how to aggregate the semantics. By designing the functions with a hierarchical structure (i.e., a context interpreter ($F_{\phi_\alpha}$) and an aggregator ($F_{\phi_\beta}$)), we let $F_\phi(\cdot)$ effectively perform two types of roles (i.e., $\phi=\{\phi_\alpha, \phi_\beta\}$).

$F_{\phi_\alpha}$ takes the set of contexts, $C_{i}$, containing the target item, $i$, as input and encodes each context, $C_{i}^{k}$. We utilize the structure of $f_{\theta_{M}}$ as $F_{\phi_\alpha}$ with our intuition that the role of $F_{\phi_\alpha}$ to interpret the contexts in which the target item is consumed by users is equivalent to extracting the current state of the user from the sequence of items consumed. When utilizing the Transformer as the context interpreter (BERT4Rec-CITIES), the description of $F_{\phi_\alpha}$ is as in Equations~(\ref{eq:bert}). When utilizing the GRU as a context interpreter (GRU4Rec-CITIES), it is only necessary to replace the self-attention layer with the GRU. The difference in doing so is that the last hidden representation vector is used as the context representation, not the center.

On top of $F_{\phi_\alpha}$, the aggregator, $F_{\phi_\beta}$, enhances each context representation by modeling dependencies among contexts using self-attention layer with parameters that are separate from $F_{\phi_\alpha}$. The enhanced representations are averaged and sent to a feed-forward layer to infer the embedding vectors. By these processes, $F_{\phi_\beta}$ can comprehensively understand the contexts by integrating multiple pieces of contextual information and infers the embeddings using the integrated representation. The inferred embedding of the target item, $e_{i}$, is computed as follows:
\begin{equation}
F_\phi(C_i) = F_{\phi_\beta}({F_{\phi_\alpha}}({C_{i}^1}), \ldots, {F_{\phi_\alpha}}({C_{i}^K})),
\end{equation}
\begin{equation}
    e_i=
    \begin{cases}
    f_{\theta_{E}}(s_{[i]}), & \text{if}\ i\in \mathcal{I}^h, \\
    F_{\phi}(C_i), & \text{otherwise}.
    \end{cases}
\end{equation}
We thus update the tail- and new-item embeddings by the embedding-inference function trained with the head items' contexts and their embeddings.

To robustly train a large number of parameters, we leverage pre-trained parameters of the sequence-modeling layer ($f_{\theta_{M}}$) in the first phase as the context interpreter ($F_{\phi_\alpha}$) and freeze the parameters. It is widely known that pre-training is more effective than learning from scratch~\cite{pretrain1, pretrain2}. Additionally, to effectively exploit the pre-trained parameters, we differently generate the training set, $C_{i}$, for learning the inference function, depending on the type of the sequential recommendation model. Each model has a different inductive bias. The inductive bias allows a learning algorithm to prioritize one solution or interpretation over another~\cite{bias}. 
BERT4Rec, which uses Transformer, assumes that some elements of the sequence can be more important depending on the relationship with other elements, and it was trained so that items appearing later can affect the target masked item. 
On the other hand, GRU4Rec, using a type of RNN as the sequence-modeling layer, assumes that the data have strong sequential dynamics and temporal dependencies, and it was trained by predicting the next item using the previous sequence of items. 
Thus, each $C_{i}^k$ is configured as follows. For BERT4Rec, we set ${\omega}_1={\omega}_2=\floor*{\frac{\ell-2}{2}}$ to incorporate the context from both directions. For GRU4Rec, we set ${\omega}_1=\ell-1$ and ${\omega}_2=0$ to incorporate the context from left to right.

\subsection{Third Phase: Applying Inferred Embeddings}
The third phase aims to produce relevance scores for each sequence by applying the inferred item embeddings to the pre-trained model of the first phase. The inferred embedding affects the item-embedding layer, $f_{\theta_{E}}$, and the recommendation layer, $f_{\theta_{R}}$, because the inferred item embedding, $F_\phi(C^i)$, is used in both layers. Reflecting the updated item embedding, we newly compute the relevance score, $r^{u}$, as follows:
\begin{equation}
r^{u}=f_{\theta_{R}}(f_{\theta_{M}}({f_{\theta'_{E}}(S^u)}^\intercal), \theta'_{E}),
\end{equation}
where $\theta'_{E}$ presents that $\theta_{E}$ is updated via the embedding-inference function.
Note that when we compute the relevance scores, we need not fine-tune the original model's parameters from the first phase. Recall that the function, $F_{\hat{\phi}}(\cdot)$, was trained to infer the original embedding of the head items. This causes the inferred embedding vectors to be distributed in a space that does not deviate significantly from the original embedding vector space. Because fine-tuning is not required, when a new item appears and is consumed only a few times by other items, it is possible to infer the embedding vector of the new item without further learning.

\section{Experiment}
In this section, we present the experimental setup and results. Our experiments were designed to answer the following research questions: 
\begin{enumerate}[label={\textbf{RQ\arabic*:}},leftmargin=*]
    \item Does \ours ~outperform baselines in terms of the recommendation performance of both the head and tail items?
    \item How can \ours ~improve the recommendation performance of the head and tail items?
    \item Are the components of \ours ~essential?
    \item What factors affect the performance of \ours?
    \item Can \ours ~infer new-item embeddings?
\end{enumerate}

\subsection{Datasets}
\begin{table}[]
\centering
\caption{Dataset statistics.}
\label{tab:statistics}
\begin{tabular}{lrrrrr}
\hline
Dataset & \#users & \#items & \makecell{avg.\\actions\\/user} & \makecell{avg.\\actions\\/item} & \#actions \\
\hline Yelp & 286,130 & 185,723 & 15.9 & 24.4 & $4.5 \mathrm{M}$ \\
Movies\&TV & 309,505 & 121,678 & 11.6 & 29.5 & $3.6 \mathrm{M}$ \\
\hline
\end{tabular}
\end{table}

We used two public datasets from real-world applications: Yelp\footnote{\url{https://www.yelp.com/dataset}} and Amazon\footnote{\url{http://deepyeti.ucsd.edu/jianmo/amazon/index.html}}. The Yelp dataset contains user reviews of local businesses, such as restaurants, dentists, and bars. The Amazon dataset contains user product-review behaviors for various categories. We selected the Movies\&TV category from this collection. To enable the implicit feedback setting, we used all observed reviews for the items as positive feedback. Then, we grouped the reviews by user and built the sequence for each user by sorting reviews according to the timestamps. To filter the noisy data, we kept users having at least five reviews. The statistics of the datasets after preprocessing are summarized in Table~\ref{tab:statistics}.

\subsection{Baselines}
We compared \ours ~with a variety of baselines, ranging from state-of-the-art sequential recommendation models to long-tail ones, to demonstrate the effectiveness of our framework. For baselines, we included some traditional methods having strong performance in previous sequential recommendation studies~\cite{gru4rec, dejavu}, and sequential recommendation methods based on BERT4Rec and GRU4Rec.

\textit{Traditional methods}:
\begin{itemize}
    \item Global Popularity (POP). This method ranks items in descending order by popularity in the training set.
    \item Sequence Popularity (S-POP). This method ranks items in descending order by popularity in the target user’s behavior sequence. Ties are broken up using POP.
    \item First-order Markov Chain (FOMC). Following the Markov assumption, this method ranks item according to transition probability, given the item in the last action, which is estimated in the training set.
\end{itemize}

\textit{Methods based on BERT4Rec}:
\begin{itemize}
    \item BERT4Rec~\cite{bert4rec}. This method uses Transformer as the sequence-modeling layer to model user behavior sequences through the ``masked-language model’’ training objective.
    \item BERT4Rec-Reranking. This method re-ranks the pre-ranked item list of BERT4Rec. Following \cite{reranking1}, the criteria for re-ranking is based on popularity from lowest to highest. The size of the pre-ranked list is set to five times that of the reranked list.
\end{itemize}

\textit{Methods based on GRU4Rec}:
\begin{itemize}
    \item GRU4Rec~\cite{gru4rec}. This method uses the GRU as the sequence-modeling layer to model user behavior sequences.
    \item GRU4Rec-Reranking. This method follows the re-ranking method used in BERT4Rec-Reranking. The only difference is that pre-ranking is given by GRU4Rec.
    \item S-DIV~\cite{sdiv}. This method clusters tail items using their content features and creates pseudo ground truth for tail items by relocating the tail cluster. The model is trained using the pseudo ground truth and listwise ranking loss function. When testing, we replaced the predicted tail cluster to an actual tail item that is closest to the centroid of the cluster. S-DIV was only applied to GRU4Rec because of the loss function.
\end{itemize}

\subsection{Implementation Details}
For common hyperparameters in all baselines, we considered the learning rate from $\left[0.0001, 0.0002, 0.001, 0.002\right]$ and the $l_2$ regularizer from $\left[0.0001, 0.0005, 0.001, 0.005\right]$. The maximum sequence length, $\ell$, was set to 50, as roughly determined by the mean number of actions per user in the previous study~\cite{sasrec}. We explored optimal hyperparameters using the validation set with the size of the item-embedding and hidden dimensions of the sequence-modeling layer. The batch size was fixed to 128. For all other hyperparameters and initialization strategies in each method, we followed the suggestions from extant works or tuned the validation set. 

\ours ~(default version) was implemented with the following settings. For pre-training sequential recommendation models of the first phase, we used BERT4Rec and GRU4Rec as baselines. During the second phase, for target embedding of the training objective, we used only the lookup matrix without the positional embedding matrix as $\theta_M$. For the aggregator, $F_{\phi_\beta}$, we set the number of self-attention blocks, $N$, to 2 and the number of heads in each block to 4. For training as the few-shot learning task, the number of contexts, $\kappa$, for the target item was randomly sampled for each iteration, and its upper limit was set to 10. For BERT4Rec-CITIES, window sizes, $\omega_1$, and $\omega_2$ were set to 24. For GRU4Rec-CITIES, $\omega_1$ and $\omega_2$ were set to 49 and 0, respectively. We trained the model using Adam optimizer~\cite{adam} with a learning-rate warmup strategy~\cite{warmup}. When we compared \ours ~with the baselines, the tail threshold, $\tau$, which is the ratio of tail items among all items, was set to 50\%. That is, half of the items belonged to the tail items based on popularity. In this setting, the tail items appeared less than 7 in the Yelp dataset and less than 6 in the Movies\&TV dataset. The source code of the proposed framework is available online\footnote{\url{https://bit.ly/32UZ2wA}}.

\subsection{Evaluation Protocols}
To evaluate the recommendation performance, we adopted the \textit{leave-one-out} evaluation protocol, which has been widely used in the literature~\cite{sasrec, bert4rec}. For each user, we treated the most recent item of the behavior sequence as test data, the second most recent item as the validation set, and the rest for training. To avoid excessive computation on all user-item pairs, we paired each ground-truth item in the test set having 100 negative items that the user had not interacted with. Following the strategy in~\cite{bert4rec}, these negative items were sampled according to probability proportional to popularity in the test data. We ranked these items with the ground-truth item. The performance of a ranked list was judged by \textit{hit ratio}@$k$ (HR@$k$) and \textit{mean reciprocal rank} (MRR). HR@$k$ computes the percentage of times the ground-truth item is among the top-$k$ items, and MRR is a position-aware metric that assigns larger weights to higher positions (i.e., $1/i$ for the $i^{\text{th}}$ position in the ranked list). Ranking the ground-truth item higher was more preferable. For S-DIV, MRR could not be computed when the ground truth belonged to the predicted cluster and was not the centroid.

\begin{table*}[]
\centering
\caption{Performance comparison of all methods in terms of head, tail, and all items.}
\label{tab:model_comparison}
\begin{tabular}{ccccccccccc}
\hline \multirow{2}{*} { Dataset } & \multirow{2}{*} { Method } & \multicolumn{3}{c} { Head items } & \multicolumn{3}{c} { Tail items } & \multicolumn{3}{c} { All items } \\
& & HR@5 & HR@10 & MRR & HR@5 & HR@10 & MRR & HR@5 & HR@10 & MRR \\
\hline \multirow{10}{*} { Yelp } & POP & 0.0589 & 0.1163 & 0.0589 & 0.0000 & 0.0000 & 0.0106 & 0.0512 & 0.1011 & 0.0526 \\
& S-POP & 0.0992 & 0.1542 & 0.0994 & 0.0286 & 0.0286 & 0.0361 & 0.0896 & 0.1374 & 0.0912 \\
& FOMC & 0.1296 & 0.1424 & 0.1087 & 0.0101 & 0.0102 & 0.0199 & 0.1140 & 0.1251 & 0.0971 \\
\cline { 2 - 11 } & BERT4Rec & \bf{0.5422} & \bf{0.7055} & \bf{0.3696} & 0.1550 & 0.2463 & 0.1087 & 0.4917 & 0.6456 & 0.3356 \\
& BERT4Rec-Reranking & 0.2352 & 0.2532 & 0.2476 & 0.0851 & 0.1473 & 0.0826 & 0.2157 & 0.2395 & 0.2262 \\
& BERT4Rec-CITIES & 0.5411 & 0.7043 & 0.3682 & \bf{0.3163} & \bf{0.4460} & \bf{0.2059} & \bf{0.5117} & \bf{0.6706} & \bf{0.3470} \\
\cline { 2 - 11 } & GRU4Rec & 0.4167 & 0.5808 & 0.2809 & 0.0359 & 0.1002 & 0.0466 & 0.3670 & 0.5180 & 0.2503 \\
& GRU4Rec-Reranking & 0.1793 & 0.2120 & 0.1847 & 0.0486 & 0.1103 & 0.0517 & 0.1623 & 0.1988 & 0.1674 \\
& S-DIV & 0.4097 & 0.5732 & - & 0.0278 & 0.0589 & - & 0.3598 & 0.5061 & - \\
& GRU4Rec-CITIES & \bf{0.4206} & \bf{0.5868} & \bf{0.2830} & \bf{0.2618} & \bf{0.3805} & \bf{0.1659} & \bf{0.3997} & \bf{0.5598} & \bf{0.2677} \\
\hline \multirow{10}{*} { Movies\&TV } & POP & 0.0595 &	0.1152 & 0.0586 & 	0.0000 & 0.0000 & 0.0103 & 0.0547 & 0.1059 & 0.0547 \\
& S-POP & 0.0633 & 0.1187 &	0.0624 & 0.0020 & 0.0020 & 0.0123 &	0.0584 & 0.1094 & 0.0584 \\
& FOMC & 0.1618 & 0.1981 & 0.1345 & 0.0136 & 0.0137 & 0.0233 & 0.1500 & 0.1833 & 0.1256 \\
\cline { 2 - 11 } & BERT4Rec & 0.3155 & 0.4289 & 0.2452 & 0.0611 & 0.0799 & 0.0614 & 0.2951 & 0.4009 & 0.2304  \\
& BERT4Rec-Reranking & 0.1907 &	0.2333& 	0.1940& 	0.0473& 	0.0653& 	0.0571 &	0.1791& 	0.2197 	& 0.1830 \\
& BERT4Rec-CITIES & \bf{0.3173} &	\bf{0.4304}& 	\bf{0.2464}& 	\bf{0.1112}& 	\bf{0.1534}& 	\bf{0.0760}& 	\bf{0.2985} &	\bf{0.4047} &	\bf{0.2327} \\
\cline { 2 - 11 } & GRU4Rec & 0.2182 &	0.3094 &	0.1781 &	0.0052 &	0.0089 	&0.0164 &	0.2011 &	0.2853 &	0.1652 
 \\
& GRU4Rec-Reranking & 0.1466 &	0.1952 &	0.1487 &	0.0060 &	0.0133 &	0.0174 &	0.1354 	&0.1806 &	0.1382 
 \\
& S-DIV & 0.1899 &	0.2818 	& -	& 0.0165 &	0.0224& 	-	&0.1760 &	0.2610& 	-\\
& GRU4Rec-CITIES & \bf{0.2201} &	\bf{0.3120}& 	\bf{0.1794} 	&\bf{0.0414} &	\bf{0.0569} &	\bf{0.0444}& 	\bf{0.2057}& 	\bf{0.2915} 	&\bf{0.1686}  \\
\hline
\end{tabular}
\end{table*}

\subsection{Experimental Results}
\subsubsection{Model Comparison}
We summarize the recommendation performance of \ours ~for all baselines in the two datasets in Table~\ref{tab:model_comparison} (\textbf{RQ1}). The best solutions in each sequential recommendation model are highlighted in bold. The performance was evaluated for each group by separating items into head and tail groups. Our methods (BERT4Rec-CITIES and GRU4Rec-CITIES) outperformed all other methods on the tail-item group in the Yelp and Amazon Movies\&TV datasets. The improvement in the tail-item group increased the overall performance shown by the performance of the all-item group. In the case of long-tail recommendation methods, S-DIV and reranking, compared with the original sequential recommendation methods, we observed some increased performance for the tail-item group, but we observed decreased performance for all of the head-item groups. The performance degradation of long-tail recommendation methods was caused by neither method explicitly improving the tail-item embeddings. In S-DIV, the process of selecting one tail item from the predicted tail cluster caused a decrease in performance, and by re-ranking methods, the reason is that the tail items were not included in the pre-ranked list. However, \ours ~explicitly improved tail-item embeddings. For the head-item group, our methods also showed comparable performance.

\begin{table}[]
\centering
\caption{Further analysis of the performance of head items.}
\label{tab:further_head}
\begin{tabular}{ccc}
\hline Dataset & Method & \makecell{Head items\\w/ tail} \\
\hline \multirow{4}{*} { Yelp } & BERT4Rec & 0.7330 \\
& BERT4Rec-CITIES & \bf{0.7337} \\
\cline { 2 - 3 } & GRU4Rec & 0.5807 \\
& GRU4Rec-CITIES & \bf{0.6001}  \\
\hline \multirow{4}{*} { Movies\&TV } & BERT4Rec & 0.3808 \\
& BERT4Rec-CITIES & \bf{0.3896} \\
\cline { 2 - 3 } & GRU4Rec & 0.2131 \\
& GRU4Rec-CITIES & \bf{0.2293} \\
\hline
\end{tabular}
\end{table}

We next ask how the performance of the head-item group can improve as the tail-item embeddings are updated (\textbf{RQ2}). We thus conducted further analysis to grasp when the performance improvement of the head-item group was achieved. We evaluated the performance of HR@10 only for the sequence wherein the ground truth was the head item, but the tail items were contained. Table \ref{tab:further_head} shows that our methods outperformed the baselines in all cases when the sequence contained tail items. These results demonstrate that improving the tail-item embeddings also helps increase the recommendation performance of the head items and the tail items. The performance improvement was more pronounced in GRU4Rec-CITIES. This is because GRU4Rec was more susceptible to situations, wherein tail items included in the sequence interfered more with the interpretation of the sequence than did BERT4Rec based on the self-attention block.

\begin{figure*}
  \centering
  \includegraphics[width=0.79\textwidth]{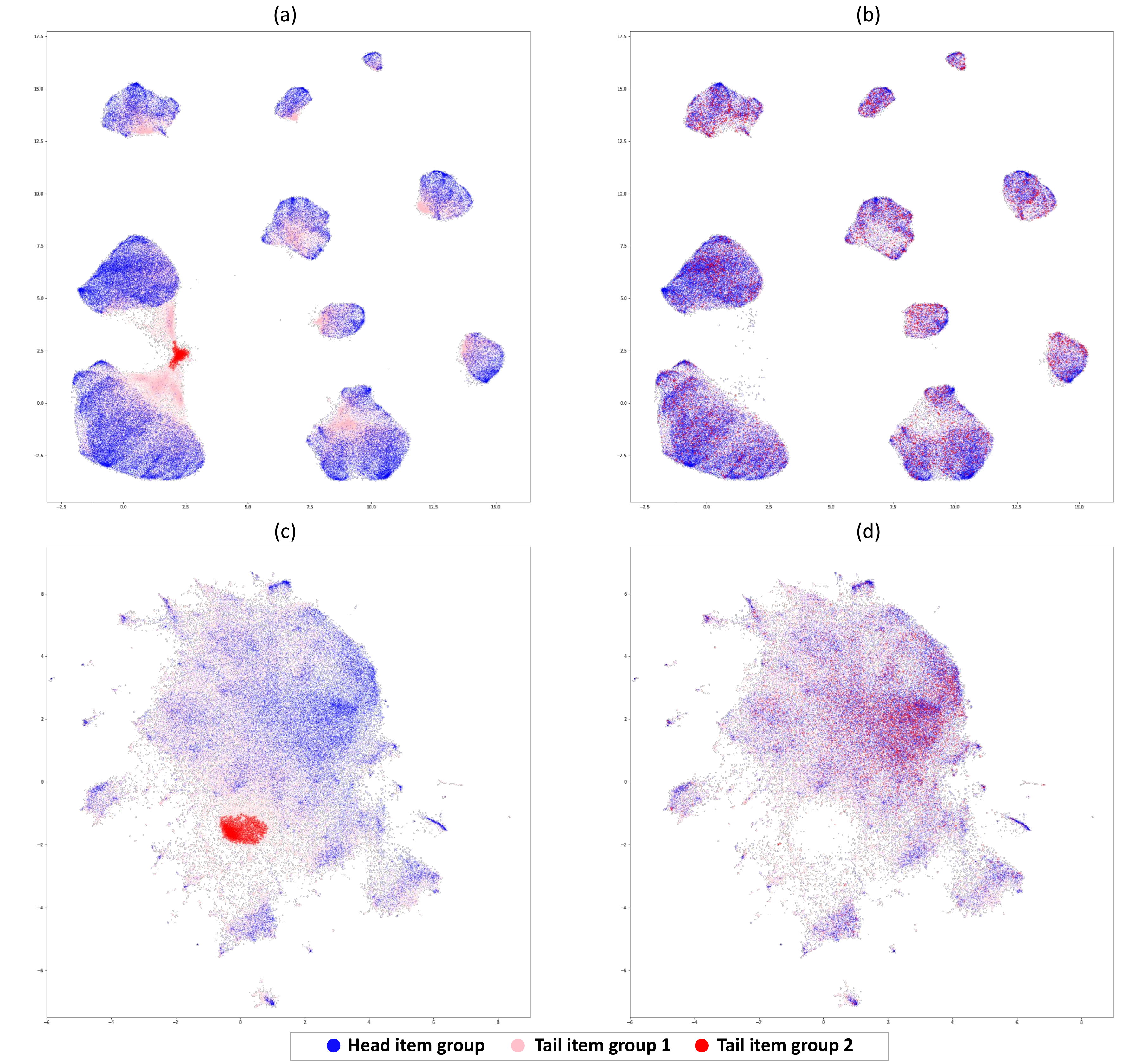}
  \caption{UMAP visualization of the item embeddings. Head items are indicated by a blue point. According to their position in the UMAP space, we split the tail items into two groups: normal tail-item group (i.e., group 1) and outlier tail-item group (i.e., group 2). The tail-item group 1 and 2 are indicated by a pink and a red point, respectively. Tail-item group 2 is farther away from the head-item group than tail-item group 1. (a) Yelp dataset using BERT4Rec. (b) Yelp dataset using BERT4Rec+CITIES. (c) Movies\&TV dataset using BERT4Rec. (d) Movies\&TV dataset using BERT4Rec+CITIES.}
  \label{fig:umap}
\end{figure*}
To answer the remaining \textbf{RQ2}, regarding how the performance improvement of the tail item can be achieved, we visualized how the inferred embeddings were distributed in vector space using UMAP~\cite{umap}. Figure~\ref{fig:umap} shows how the tail-item embeddings were updated (from left to right) by \ours ~using the two datasets. Originally, most tail items were distributed far enough to be distinguished from the head-item group. However, the inferred tail items by \ours ~spread widely into space where the head items were distributed. This happened with both datasets, and it stands out for the tail-item group 2 indicated by the red point. In the case of BERT4Rec in the Yelp dataset, HR@10 of the items in tail-item group 2 was only 0.0289. When \ours ~was applied, HR@10 was 0.4362 for the same items (1,409\% increased). This was also the case with the Movies\&TV dataset using BERT4Rec. HR@10 of the items in tail-item group 2 increased from 0.0184 to 0.0432 (135\% increased). Note that the items in tail-item group 2 were distributed more densely than were the other items. We can infer that this is because the items that were rarely learned did not deviate significantly from their initial parameters. In fact, the average number of reviews of the items in tail-item group 2 was only 1.2 for both Yelp and Movies\&TV datasets: significantly less than the average number of reviews of other tail items. Even with these items, using our framework, the items were not densely distributed in one area but were instead spread widely throughout. 

\subsubsection{Ablation Study}
\begin{table}[]
\centering
\label{tab:ablation}
\caption{Ablation analysis of BERT4Rec-CITIES on two datasets.}
\begin{tabular}{cc lll}
\hline Dataset & Setting & Head & Tail & All \\
\hline \multirow{5}{*} { Yelp } & default & 
0.7043 & 	\bf{0.4460} & 	\bf{0.6706}  \\
& Use $\mathcal{I}$ for target item & {0.7061} & 0.2322$\downarrow$ & 0.6443\\
& w/o few-shot learning & 0.7025 &0.4357 &  0.6677\\
& w/o pre-training $\phi_\alpha$ & 0.7051 & 0.4168$\downarrow$ & 0.6675 \\
& w/o freezing $\phi_\alpha$ &  0.7049 & 	0.4342 	& 0.6695 \\
\hline \multirow{5}{*} { Movies\&TV } & default & 
\bf{0.4304} &	\bf{0.1534} &	\bf{0.4047}  \\
& Use $\mathcal{I}$ for target item & 0.4289 & 0.0566$\downarrow$ & 0.3990\\
& w/o few-shot learning &  0.4290 &	0.1054$\downarrow$ & 0.4030\\
& w/o pre-training $\phi_\alpha$ & 0.4301 &	0.0893$\downarrow$ &	0.4028  \\
& w/o freezing $\phi_\alpha$ & 0.4302 	&0.1381 &	0.4035 \\
\hline
\end{tabular}
\end{table}

To answer \textbf{RQ3}, we performed an ablation study over key components of \ours ~to analyze their impacts. We present four variants of our default setting as follows:
\begin{itemize}
    \item Use $\mathcal{I}$ for target item: As opposed to using only the head items for the target in our default setting, when training the embedding-inference function, $F_{\phi}$. This setting also uses the tail items for the target. 
    \item Without few-shot learning: When training $F_{\phi}$, we take all contexts of the target item as input in this setting, unlike the default setting, which takes $\kappa$ contexts of the target item.
    \item Without pre-training $\phi_\alpha$: We train the context interpreter, $F_{\phi_\alpha}$, from scratch without utilizing the parameters of $f_{\theta_{M}}$ in the pre-trained model.
    \item Without freezing $\phi_\alpha$: We utilize the parameters of $f_{\theta_{M}}$ as $F_{\phi_\alpha}$, but we do not freeze it.   
\end{itemize}
Table IV shows the performance of our default setting and the variants based on HR@10. The best solutions are highlighted in bold and $\downarrow$ indicates performance drop more than 5\%. The most significant performance degradation occurred when we use $\mathcal{I}$ for the target item. This implies that training $F_{\phi}$ using only high-quality head-item embeddings as the target was more effective because of the low-quality of tail-item embeddings. The performance degradation without few-shot learning implies that training our function with $\kappa$ contextual items was essential to inferring tail-item embeddings with few contextual items. Additionally, the results of without pre-training and without freezing settings imply that our pre-training strategy was effective. $f_{\theta_{M}}$, leveraged from the pre-trained model, helped us train $F_{\phi}$, which had a hierarchical structure. We can thus infer that the roles of $f_{\theta_{M}}$ and $F_{\phi_\alpha}$ are similar.

\subsubsection{Impact of Tail threshold $\tau$}
\begin{figure}
  \centering
  \includegraphics[width=85mm]{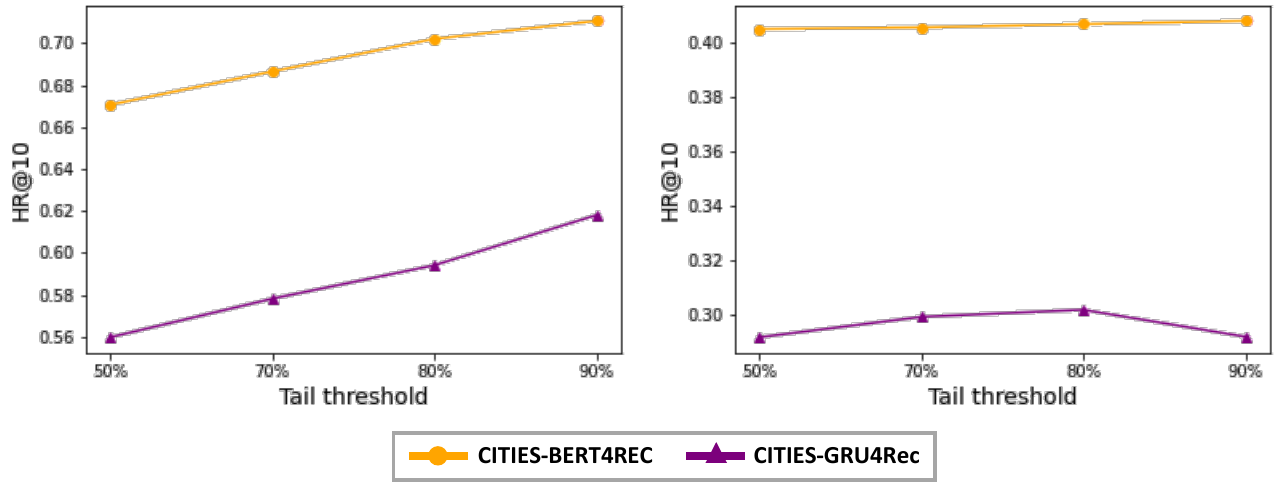}
  \caption{HR@10 of our methods according to the tail thresholds (left) Yelp and (right) Movies\&TV datasets.}
  \label{fig:tt}
\end{figure}
To answer \textbf{RQ4}, we first investigated the effect of tail threshold $\tau$. It is used to determine whether the item embeddings are inferred by \ours. As $\tau$ increases, a larger number of items are considered as tail-item groups, and their embeddings are inferred. This reduces the amount of training data for our embedding-inference function because only the head items are used as the ground truth when training. However, the quality of the ground-truth items relatively improve. Figure~\ref{fig:tt} shows the performance change for the all item group according to the tail threshold. As the tail threshold increases, the performance tends to improve. This implies that, if the quality of the training data is fine, the embedding-inference function can be robustly trained using a small amount of data. However, a large threshold does not always lead to better performance. This fact is evident when examining the Movies\&TV dataset having fewer items than the Yelp dataset. This explains that there was a trade-off between data quantity and quality when training the embedding-inference function.

\subsubsection{Impact of the Number of Contexts, $K$}
\begin{figure}
  \centering
  \includegraphics[width=85mm]{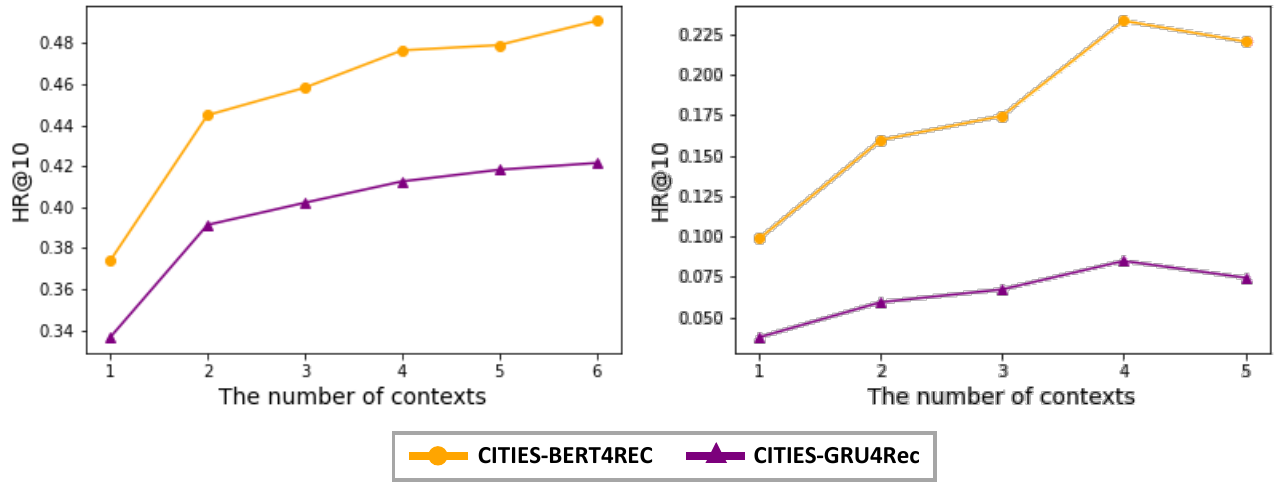}
  \caption{HR@10 of our methods according to the number of contexts for target tail item (left) Yelp and (right) Movies\&TV datasets.}
  \label{fig:noc}
\end{figure}
Next, we explain our findings on the effect of the number of contexts, $\kappa$, on the performance in response to \textbf{RQ4}. We stated that the embedding-inference function requires the ability to infer embeddings in a few contexts because the number of times, $\kappa$, the tail item is consumed is relatively few. Figure~\ref{fig:noc} shows the performance change for the all-item group according to $\kappa$. The most interesting observation is that the performance tended to converge as $\kappa$ increased. This demonstrates that, by the few-shot learning formulation, the embedding-inference function was trained to work properly, even in situations where there was insufficient context information in the tail item. A larger $\kappa$ does not always lead to better performance because the diversity of context in which items are consumed varies from item to item. To understand the meaning of items consumed in various contexts, more evidence is required. The range of the x-axis is different in the two plots, because the criteria of the frequency values belonging to the tail-item group are different in the two datasets, despite having same tail threshold.

\subsubsection{Results of New Items}
We conducted an additional experiment to verify that \ours ~can infer the new-item embeddings (\textbf{RQ5}) without further training. We evaluated the performance (HR@10) by separating the items into the extant and new groups. For this, we split the data used to learn \ours ~from the previous experiment into two based on the time of occurrence. We used only half the amount that occurred earlier to learn the embedding-inference function in this experiment. When the training was completed, the other half was used to infer the embedding of new items. This setting was intended to prevent \ours ~from seeing contexts where the new items appeared during training.
Table~\ref{tab:new_items} compares the performance of the extant and new items with HR@10. It shows that, in all cases, the performance of the new tail items was comparable to that of the extant tail items. We can thus infer that the slightly lower performance in the new-item group was caused by the patterns of context that were not seen during training but appeared when inferring the embedding of the new item. However, as users increasingly consume new items over time, 
by leveraging the contexts of that consumption, \ours ~can infer new-item embeddings of better quality than extant tail-item embeddings. This can be confirmed from the result, which showed that the performance of the head item of the new-item group was higher than that of the tail item of the extant-item group. Moreover, considering that the models not applying \ours ~could not handle new items without using content features, it can be said that \ours ~inferred new-item embeddings as high quality similar to the extant tail-item embeddings.
\begin{table}[]
\centering
\caption{Performance of \ours ~for the new items compared to the extant items}
\label{tab:new_items}
\addtolength{\tabcolsep}{-3pt}
\begin{tabular}{cccccc}
\hline \multirow{2}{*} { Dataset } & \multirow{2}{*} { Method } & \multicolumn{2}{c} { Extant items }  & \multicolumn{2}{c} { New Items }  \\ & & Head & Tail & Head & Tail \\
\hline \multirow{2}{*} { Yelp } & BERT4Rec-CITIES & 0.7046 & 0.5145 & 0.5764 & 0.4497 \\
& GRU4Rec-CITIES & 0.5880 & 0.4244 & 0.4556  & 0.3624 \\
\hline \multirow{2}{*} { Movies\&TV } & BERT4Rec-CITIES & 0.4383 & 0.1185 & 0.1207 & 0.0840 \\
& GRU4Rec-CITIES & 0.3130 & 0.0656 & 0.0692 & 0.0491\\
\hline
\end{tabular}
\end{table}

\section{Conclusion}
In this paper, we proposed \ours ~after scrutinizing the structural reason for why tail items are barely served in general sequential recommendation models. Our framework precisely inferred tail-item embeddings by training the embedding-inference function to grasp the meaning of the item in the context in which it was consumed. We found that, by applying \ours ~to state-of-the-art methods, we improved the recommendation performance not only for the tail items but also for the head items of two real-world datasets. Moreover, we found that \ours ~could infer embeddings of the new items without the further learning process. 

Two promising directions for future research remain. The first direction is to utilize user information to infer embedding. Even if the item was consumed in the same context, the meaning could be subtly different depending on which user consumes. The second direction is to iteratively perform the general sequential recommendation learning method and our embedding-inference method for the tail item until convergence. The improved tail-item embeddings via the latter method can help model sequential dynamics in the former method, which in turn improves head-item embeddings.

\section*{Acknowledgement}
The authors are grateful to Vinnam Kim, Jinbae Im, Gyeongbok Lee, and Sunyou Lee for their helpful comments.

\bibliographystyle{IEEEtran}
\bibliography{IEEEabrv, main}
\end{document}